\documentclass[pra,twocolumn,amsmath,amssymb,superscriptaddress,showpacs]{revtex4}
\usepackage{graphics}
\usepackage{epsfig}
\usepackage{bbm}
\usepackage[latin1]{inputenc} 

\newcommand{\be}{\begin{equation}}
\newcommand{\ee}{\end{equation}}
\newcommand{\eea}{\end{eqnarray}}
\newcommand{\bea}{\begin{eqnarray}}

\newcommand{\mean}[1]{\ensuremath{\langle{#1}\rangle}}
\newcommand{\norm}[1]{\ensuremath{\Vert{#1}\Vert}}

\newcommand{\eins}{\ensuremath{\mathbbm 1}}
\newcommand{\qed}{\ensuremath{\hfill \Box}}

\newcommand{\fh}{\ensuremath{\mathfrak{h}}}

\newcommand{\II}{\ensuremath{\mathcal{I}}}
\newcommand{\RR}{\ensuremath{\mathcal{R}}}
\newcommand{\NN}{\ensuremath{\mathcal{N}}}

\newcommand{\ketbra}[1]{\ensuremath{| #1 \rangle \langle #1 |}}

\newcommand{\ket}[1]{\ensuremath{|#1\rangle}}

\newcommand{\bra}[1]{\ensuremath{\langle#1|}}

\newcommand{\kommentar}[1]{}
\renewcommand{\vr}{\ensuremath{\varrho}}

\begin{document}
\title{Energy and multipartite entanglement in multidimensional 
and frustrated spin models}
\date{\today}
\begin{abstract}
We investigate the relation between the 
entanglement properties of a quantum state
and its energy for macroscopic spin models. 
To this aim, we develop a general method to compute 
energy bounds for states without certain forms of multipartite 
entanglement. Violation of these bounds implies 
the presence of these types of multipartite entanglement.
As examples, we investigate the Heisenberg model
in different dimensions, the Ising model and the 
XX model in the presence of a magnetic field.  Finally,
by studying the Heisenberg model on a triangular lattice, 
we demonstrate that our techniques can be applied also to 
frustrated systems.
\end{abstract}

\author{Otfried G\"uhne}

\affiliation{Institut f\"ur Quantenoptik und Quanteninformation,
\"Osterreichische Akademie der Wissenschaften,
A-6020 Innsbruck, Austria}

\author{G\'eza T\'oth}

\affiliation{Max-Planck-Institut f\"ur
Quantenoptik, Hans-Kopfermann-Stra{\ss}e 1,
 D-85748 Garching, Germany}

\affiliation{Research Institute of Solid State Physics and Optics,
Hungarian Academy of Sciences, H-1525 Budapest P.O. Box 49, Hungary}

\pacs{03.65.-w, 03.67.-a, 05.30.-d}

\maketitle

\section{Introduction}

In recent years, the investigation of entanglement 
in condensed matter systems has become one of the main
lines of research in quantum information science 
\cite{allcondqiv1,allcondqiv2,allcondqiv3}. 
The increased interest in this topic is fed by 
several motivations. On the one hand, the studies helped 
to understand fundamental properties of condensed matter 
systems like quantum phase  transitions. On the other hand, 
they lead to results of practical importance
since they allowed to design new simulation techniques 
for the calculation of ground state energies of spin models 
\cite{vidal}.

What kinds of entanglement occur in natural situations?
This question provides another motivation for studying 
entanglement properties of condensed matter systems. 
Indeed, condensed matter systems and especially spin 
models are natural candidates for our studies
where various forms of entanglement might occur, 
mainly at low temperatures.
One possibility to study the presence of entanglement 
in spin systems is to relate the energy or other 
macroscopic observables of the system  to certain 
entanglement properties of the state \cite{geza, vlatko, 
dowling, allhamiltonian, allrest, us}. 

In this paper we attempt to proceed in this direction 
by investigating the relation between the energy of a state 
and its {\it multipartite} entanglement properties.
We will derive a general method for calculating energy 
thresholds for states without certain types of multipartite
entanglement. Below these energies, and consequently below 
a certain temperature, the state  must therefore contain 
multipartite entanglement. Our approach is, however, not 
restricted to states in thermal equilibrium. We demonstrate 
that our method can successfully be applied to various models 
and also to frustrated systems. In this way, we extend the 
results of Ref.~\cite{us} where such energy thresholds have been 
computed for two special spin models in one dimension. 

Our paper is divided into four sections. These are 
organized as follows. In Section II we introduce the 
notion of multipartite entanglement that we use in 
this paper. That is, we explain the definition of 
$k$-producibility. 
We also pose the problem that we want to solve. In Section~III 
we present our method for computing the desired energy 
bounds. We present in detail the calculation for a 
two-dimensional Heisenberg model on a square lattice, 
the generalization to other models is then straightforward.
In Section~IV we discuss three simple applications: the Heisenberg
model in various dimensions, and the Ising model and the XX model
with a magnetic field in one dimension. For the Ising model,
we also discuss the impact of phase transitions on our 
energy thresholds.
In Section~V we consider
the Heisenberg model a two-dimensional triangular lattice. We 
show that with some modifications our methods can also be used
to investigate multipartite entanglement in such a frustrated 
system.

\section{Definitions and statement of the problem}

Let us first explain the notion of multipartite entanglement
that we use for our study. This is the so-called $k$-producibility,
introduced in Ref.~\cite{us}. It is defined as follows:
For a pure state $\ket{\psi}$ on $N$ qubits we ask whether it 
is possible to write
\be
\ket{\psi} = 
\ket{\phi_1} \otimes \ket{\phi_2} \otimes ... \otimes \ket{\phi_K},
\label{kp1}
\ee
where the $\ket{\phi_i}$ are states of maximally $k$ qubits. If 
this is the case, then only $k$-qubit entanglement is necessary 
to generate $\ket{\psi}$ and the state $\ket{\psi}$ does not 
contain any $(k+1)$-qubit entanglement. If Eq.~(\ref{kp1}) holds, 
we call the state $k$-producible, if not, we say that $\ket{\psi}$
contains $(k+1)$-partite entanglement. Examples of four- and 
two-producible states are shown in Figs.~1 and 2.

\begin{figure}[t]
\centerline{\epsfxsize=0.9\columnwidth
\epsffile{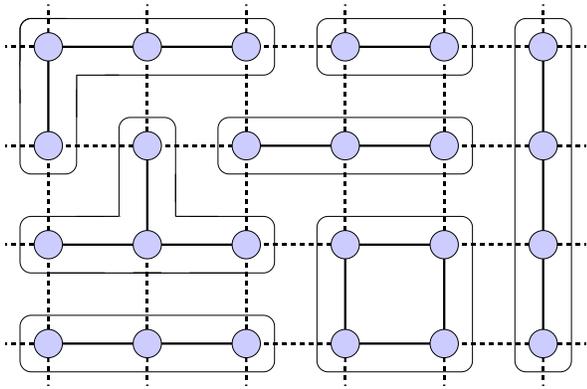} }
\caption{Schematic view of a four-producible state in a spin model 
of  24 qubits. Dashed lines correspond to interactions between 
disentangled qubits and solid lines represent the interactions 
between qubits which are allowed to be entangled.
}
\end{figure}

For mixed states, we can extend this definition via considering
convex combinations. I.e., we call a mixed state $\vr$ 
$k$-producible if we can write
\be
\vr=\sum_i p_i \ketbra{\psi_i}
\ee
with $p_i \geq 0, \sum_i p_i=1$ and $k$-producible 
$\ket{\psi_i}.$ If this is not the case, $\vr$ contains 
$(k+1)$-partite entanglement. Physically, a $k$-producible 
mixed state requires $k$-qubit entanglement and mixing for 
its creation only. Conversely, a state contains $k+1$-party
entanglement if and only if the quantum correlations of this 
state cannot be explained by assuming $k$-qubit entanglement only. 

This classification of multipartite entanglement has some 
connections to the usual notion of $k$-separability, which 
is often used for small numbers of qubits. This issue has
been discussed in Ref.~\cite{us}. Here, we only want to point 
out that the $N$-separable (fully separable) states are by 
definition the states which can be written as 
$\vr=\sum_i p_i \vr_1 \otimes ...\otimes \vr_N.$
These states are just the one-producible states.

The notion of $k$-producibility leads to a discrete classification 
of multipartite states. For pure states, it is easy to see that 
the $k$-producible states form a set of measure zero in the set
of $2k$-producible states and that in the vicinity of any 
$k$-producible state one can find states with arbitrary high 
producibility \cite{bemerkung}. For mixed states, however, 
this is not true anymore, and one can show (as for the notion 
of $k$-separability \cite{abls}) that the set of mixed 
$k$-producible states is not of measure zero in the space of 
all mixed states.

Finally, it is worth noting 
that the notion of $k$-producibility for pure 
states has a close relation to the  Schmidt measure, which is
an entanglement monotone for multi-qubit states \cite{EB04}. 
The Schmidt number of a pure multi-qubit state is defined as 
follows. One expands $\ket{\psi}$ as the sum of tensor products 
of single qubit states 
$ 
\ket{\psi} = \sum_{k=1}^R \ket{\phi_k^{(1)}} 
\otimes \ket{\phi_k^{(2)}} 
\otimes \ket{\phi_k^{(3)}} \otimes ... 
$ 
For every quantum state we take the expansion with the minimal
$R,$ which will be denoted by $r.$ Then $\log_2 (r)$ is the Schmidt
measure of  $\ket{\psi}$ and for N-qubit quantum states we have 
always $ \log_2 (r_q) \le N.$

Since the Schmidt number for two-qubit states in maximally two
and for three-qubit states is maximally three \cite{EB04}, we can 
conclude that for one-, two- and three-producible states  
\begin{eqnarray} 
\log_2 r_{1p} &=& 0,
\nonumber
\\ 
\log_2 r_{2p} 
&\le& 
\frac{N \log_2(2)}{2}
=\frac{N}{2},
\nonumber
\\ 
\log_2 r_{3p} &\le& 
\frac{N \log_2(3)}{3}
\approx 
0.53N .
\end{eqnarray} 
holds.

Now we can state the main problem we want to study in this 
paper. Let us assume that we have a macroscopic spin system 
of qubits on some lattice, which interact via some Hamiltonian
\begin{eqnarray}
H&=& \sum_{<i,j>} h_{ij},
\end{eqnarray}
which is a sum of two-qubit interactions. We always assume 
periodic boundary conditions. For this situation, we want
to derive lower bounds for $\mean{H}$ for $k$-producible 
states. That is, we want to compute a constant $E_{kp}$ such 
that 
\be
\mean{H}\geq E_{kp}
\ee
holds for all $k$-producible states. If this bound is then 
violated at low temperatures, the state under consideration 
contains $(k+1)$-party entanglement. Note, however, that we 
do not restrict our attention to states in thermal equilibrium.
Since we assume that the number of qubits $N$ is large, it 
will be convenient to express $E_{kp}$ as a rescaled energy 
per interaction bond. 

In the next Section, we will present the main idea of our 
method to compute $E_{kp}.$

\section{Estimating the energy}

\begin{figure}[t]
\setlength{\unitlength}{0.05625\columnwidth}
\begin{picture}(16,11)
\thinlines
\put(0,0){\includegraphics[width=0.9\columnwidth]{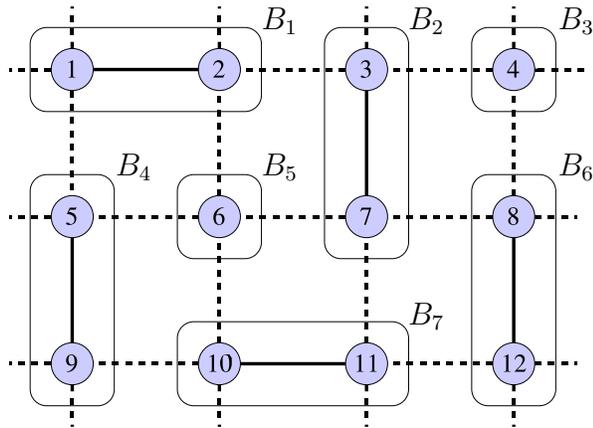}}
\put(7,11){\mbox{\large $B_1$}}
\put(11,11){\mbox{\large $B_2$}}
\put(15.1,11){\mbox{\large $B_3$}}
\put(3,7){\mbox{\large $B_4$}}
\put(7,7){\mbox{\large $B_5$}}
\put(15.1,7){\mbox{\large $B_6$}}
\put(11,2.9){\mbox{\large $B_7$}}
\end{picture}
\caption{A possible grouping for a pure two-producible
state in twelve-qubit spin system.
See text for details.
}
\label{fig_settings}
\end{figure}

In this Section, we present as the main result of the 
paper a general method to estimate the energy for 
$k$-producible states. While the main result is 
quite simple, its proof requires some technical 
effort. The logical structure is 
as follows: For a pure $k$-producible state some expectation
values in the Hamiltonian factorize and some not. 
We  collect all the factorizing terms in the Hamiltonian, 
and estimate them via the Cauchy-Schwarz inequality.
Then, we arrive at Eqs.~(\ref{cibound}, \ref{hbound}).
If we can perform the maximization in Eq.~(\ref{cibound})
then the Eq.~(\ref{hbound}) delivers the desired energy bound. 
The point is that even for macroscopic  $k$-producible 
states the maximization requires only a maximization over
$k$-qubit states. Sometimes, this can be done analytically, 
otherwise it can be solved numerically in an efficient 
manner. Finally we discuss whether the derived bounds 
are sharp.
 
So let us explain our method in the following example. 
We want to derive a bound for two-producible states 
for a two-dimensional Heisenberg lattice. That is, we 
consider the Hamiltonian
\begin{eqnarray}
H_H &=& \sum_{<i,j>} h_{ij},
\nonumber 
\\
h_{ij}&=& 
X_{i}X_{j}+
Y_{i}Y_{j}+
Z_{i}Z_{j}.
\label{XXXmodel}
\end{eqnarray}
for a two-dimensional system with periodic boundary 
conditions. Here and in the following, $X_i, Y_i, Z_i$
denote the Pauli matrices $\sigma_x^{(i)}, \sigma_y^{(i)}, 
\sigma_z^{(i)},$ acting on the $i$-th qubit. We will
exemplify the definitions required for our 
method using the example of a special two-producible 
state of twelve qubits given in  Fig.~2.
 
In order to bound the energy for two-producible states, 
it suffices to consider a generic {\it pure} two-producible 
state $\ket{\psi}.$ This comes from the fact that the 
mixed $k$-producible states form a convex set in the state
space and the pure $k$-producible states are its extremal 
points. Thus, any linear function takes its maximum in a 
pure state as an extremal point.

A fixed two-producible $\ket{\psi}$ 
results in a partition of the whole spin system into 
several one- and two-qubit blocks. Indeed, one can 
identify some pairs $i,j$ of qubits where the reduced 
state is allowed to be entangled, and some single qubits
$k,$ which are not entangled with any other qubit. 
Let us denote the total number of blocks by $K,$  
the number of one-qubit blocks by $L_1$ and the number 
of two-qubit blocks by $L_2.$ A possible blocking is 
shown in Fig.~2: We have seven blocks $B_1, ..., B_7,$ 
where $B_3$ and $B_5$ are single-qubit blocks and the 
rest are two-qubit blocks. Thus we have $K=7, L_1=2,$ 
and $L_2=5.$ It is important to note that we can restrict 
our attention to the case where the two-qubit blocks are 
between interacting qubits. This is true for the following 
reason: If a two-qubit block consists of two noninteracting 
qubits $i,j,$ then the Hamiltonian is only sensitive to 
the reduced density matrices at each qubit, i.e., it does 
only take then local properties into account. Thus, we can 
replace this two-qubit block by two one-qubit blocks.

In general, the mean value of the Hamiltonian consists 
of two-qubit expectation values of the type 
$W_{ij}=\mean{A_i A_j}, A=X,Y,Z.$ For simplicity, we denote 
in the following $\mean{A_i B_j}=a_i b_j$ and $\mean{A_i}=a_i$ 
for $A, B = X,Y,Z.$ Note that this definition implies that 
in general $a_i b_j \neq a_i \cdot b_j$ and these notations
have to be distinguished. Due to the special partition, however, 
some of the mean values factorize. For instance, in the situation 
of Fig.~2 we have $x_2 x_3= x_2 \cdot x_3.$ Now we have to find 
an efficient way for taking all these contributions into account. 

For this purpose, we  define for each block $B_i$ three 
sets of indices: $I(i)$ are the qubits which lie inside 
$B_i,$ $R(i)$ are the qubits inside $B_i$ which interact
via the Hamiltonian with qubits outside of $B_i$ and $N(i)$ 
are the qubits outside $B_i,$ which nevertheless interact 
with some qubit of $I(i).$ For instance, in the example of 
Fig.~2 we have $I(1)=\{1,2\},$ $R(1)=\{1,2\}$ and 
$N(1)=\{3,4,5,6,9,10\}.$ For our special case of a 
two-producible state we have $I(i)=R(i),$ but in general 
this does not have to be the case.
 
Now we can define for each block a set of expectation 
values from  the Hamiltonian in the following way. 
We define
\be
\II(i) := \{W_{kl} \;\;\vert\;\; k,l \in I(i)  \}
\ee
as the contributions of the Hamiltonian inside the block 
$B_i.$ In our example, this would be 
$\II(1)=\{x_1x_2, y_1y_2, z_1 z_2\}.$
Note that the set $\II(i)$ is empty for the one-qubit blocks.
In the following, we will  denote the single elements of sets
like $\II(i)$ by $\II(i)[j].$ 

Then, we collect the ``outgoing'' contributions from a block
via all the two-qubit Hamiltonians. That is, we define: 
\be
\RR(i) := 
\{(W_{kl})\vert_k \;\;\vert\;\; k \in R(i), l\in N(i)\}.
\ee
Here, using $\vert_k$ with $k \in R(i)$ we express that 
the two-qubit contributions are restricted to the operator
acting only on the qubit belonging to $R(i).$ For example, 
from a term of the form  $W_{23}=\mean{X_2 X_3}$ we take only the term 
$\mean{X_2} = x_2.$ Also, we take all terms with their 
respective multiplicity, i.e., if the qubit $k\in R(i)$ 
interacts with several qubits in $N(i),$ the same term 
appears several times in $\RR(i).$ For our example in 
Fig.~2 we would have: 
$\RR(1)=
\{
x_1, y_1, z_1, x_1, y_1, z_1,  x_1, y_1, z_1, 
x_2, y_2, z_2, x_2, y_2, z_2, x_2, 
y_2, 
\\
z_2
\}.$

Finally, we count for each block the contributions in the 
neighborhood via:
\be
\NN(i) := \{ (W_{kl})\vert_l, \;\;\vert\;\; k \in R(i), l\in N(i)\}.
\ee
These are, in a certain sense, the complementary contributions 
to the contributions in $\RR(i).$ We always write them in the 
same 
order as the contributions in $\RR(i),$ i.e. the first 
element of $\RR(i)$ should correspond to the first element of 
$\NN(i)$ in the Hamiltonian, etc. In our example, we would 
have:
$\NN(1)=
\{
x_9, y_9, z_9, x_4, y_4, z_4,  x_5, y_5, z_5, 
x_{10}, y_{10}, z_{10}, x_3, y_3, z_3, x_6, 
\\
y_6, 
z_6
\}.$
The idea behind these definitions of $\RR(i)$ and $\NN(i)$ 
is the following: The terms in $\RR(i)$ and $\NN(i)$ are 
just the ones which factorize in the Hamiltonian. Thus, 
viewing  $\RR(i)$ and $\NN(i)$ as vectors, 
the scalar product corresponds to the mean value of some 
terms in the Hamiltonian, 
$\RR(i) \cdot \NN(i) = \sum_k \RR(i)[k] \;\; \NN(i)[k] = 
\sum_{ k\in B(i), l\notin B(i)} \mean{W_{kl}}.$

To estimate $\mean{H}$ for a given two-producible state 
$\ket{\psi}$ we interpret $\RR(i)$ and $\NN(i)$ 
as real vectors. We then define 
\bea
\vec{v}_1 &:=&
\frac{1}{\sqrt{2}}\cdot\RR(1)\oplus ...
\oplus\frac{1}{\sqrt{2}}\cdot \RR(K),
\nonumber
\\
\vec{v}_2 &:=&
\frac{1}{\sqrt{2}}\cdot\NN(1)\oplus ...
\oplus\frac{1}{\sqrt{2}}\cdot \NN(K).
\eea
Please note that a term of 
the type $W_{kl}=x_k x_l = x_k \cdot x_l$ originating 
from an interaction between two blocks $B_i$ and $B_j$ 
appears {\it twice} in each of these vectors: 
one time with $x_k \in \RR(i)$ and 
$x_l \in \NN(j)$ and one time with $x_k \in \NN(i)$ and 
$x_l \in \RR(j).$ Thus, $\vec{v}_1$ and $\vec{v}_2$ are 
built of the same terms, but in different order. This
implies that $\norm{\vec{v}_1} = \norm{\vec{v}_2}.$

With this definition, it follows that 
\be
\mean{H} = \sum_{i=1}^K \sum_k \II(i)[k] + 
\vec{v}_1 \cdot \vec{v}_2 
\ee 
holds. This implies due to the Cauchy-Schwarz 
inequality that
\bea
\mean{H} &\geq& \sum_{i=1}^K \sum_k \II(i)[k] - 
\norm{\vec{v}_1}\norm{\vec{v}_2}
\nonumber
\\
&=&- \Big( 
\sum_{i=1}^K \sum_k - \II(i)[k] + \norm{\vec{v}_1}^2 
\Big).
\label{csu}
\eea
The key point is that the right hand side of this inequality
can be estimated by 
minimization
for each of the blocks 
$B_i$  separately. 
Indeed, if we define for the block $B_i$  
\be
C_i :=  \max_{\ket{\psi}} 
\big[
\sum_k - \II(i)[k] + \frac{1}{2} \sum_k (\RR(i)[k])^2
\big],
\label{cibound}
\ee
where $\ket{\psi}$ is a quantum state on the block $B_i,$ 
we have 
\be
\mean{H} \geq - \sum_i C_i.
\label{hbound}
\ee
The estimation of the $C_i$ does now {\it only} depend 
on the fact whether the block $B_i$ is a one- or a two-qubit 
block and not on the relations between these blocks. For 
the Heisenberg interaction, we have
\be
C_i = \max_{\ket{\psi}} \big[ 2 (x_k^2 + y_k^2 + z_k^2) \big] = 2
\label{c1bound}
\ee
for a one-qubit block  $B_i$ on the qubit $k$ and
\bea
C_i &=& \max_{\ket{\psi}} 
\big[
-x_k x_l - y_k y_l - z_k z_l +
\nonumber 
\\
&&
+\frac{3}{2} (x_k^2 + y_k^2 + z_k^2 + x_l^2 + y_l^2 + z_l^2)
\big]
= \frac{13}{3}
\;\;\;\;
\label{c2bound}
\eea
for a two-qubit block on the qubits $k$ and $l.$ This bound can 
be obtained from the representation of 
$\ketbra{\psi}=\sum_{k,l=1,x,y,z} \lambda_{kl} \sigma_k \otimes \sigma_l$
\cite{us}. It is also a special case of a general bound presented 
as Lemma 1 in the Appendix.

With these bounds, we immediately get for our example in 
Fig.~2 the bound 
$\mean{H_H} \geq  - (5 \cdot 13/3 + 2 \cdot 2 ) = - 77/3.$ 
For  the general case of $N$ qubits, we get
\be
\mean{H_H}  \geq  - 
\max_{L_1 +  2 L_2 = N}[2 \cdot L_1 + \frac{13}{3} \cdot L_2] 
\geq - \frac{13 N}{6} \approx - 2.16 N.
\ee
Since a two-dimensional lattice of $N$ qubits has $2N$ bonds, 
the energy per bond for two-producible states is bounded from below 
by 
\be
\frac{\mean{H_H}}{2N} \geq E_{2p} = - \frac{13}{12}.
\label{e2bound}
\ee

Two questions arise at this point. First, we have to ask 
whether this bound is useful, in the sense that it is 
violated at low temperatures. This is the case since for 
the ground state the energy per bond is $E_0= -1.338$ 
\cite{dowling,values}. Thus, in a considerable temperature 
regime, the thermal states cannot be two-producible. 

Second, the question arises whether the derived bound is 
sharp. This question deserves some discussion. The idea
to show sharpness of an obtained bound is the following: 
Let us assume we have found states $\ket{\phi_i}$ for 
which the maxima $C_i$ in Eq.~(\ref{cibound}) are obtained. 
Then we have to build out of these states $\ket{\phi_i}$
the total state $\ket{\psi}$ such that $\ket{\psi}$ saturates
the bound in Eq.~(\ref{hbound}). To do so, we have to assure, 
that for the state $\ket{\psi}$ the Cauchy-Schwarz inequality 
in Eq.~(\ref{csu}) was sharp, i.e. 
$\vec{v}_1 \cdot \vec{v}_2 = -  \norm{\vec{v}_1} \norm{\vec{v}_2}.$
This can be done in two steps: First we guarantee that
$\vec{v}_1 \cdot \vec{v}_2 = \norm{\vec{v}_1} \norm{\vec{v}_2}.$
Then, by applying some unitary transformations on the 
$\ket{\phi_i}$ we make sure that
$\vec{v}_1 \cdot \vec{v}_2 = - \norm{\vec{v}_1} \norm{\vec{v}_2}.$

Let us show how this works in our example of two-producible states
\cite{remark1}. 
Let $\ket{\phi}_{k,l}$ be the state saturating Eq.~(\ref{cibound})
on the qubits $k,l.$  Let us enumerate the qubits
as in Fig.~2. and consider the total state
$\ket{\tilde \psi}= \ket{\phi}_{1,2}\otimes \ket{\phi'}_{3,4} 
\otimes \ket{\phi}_{5,6} \otimes \ket{\phi'}_{7,8} 
\otimes \ket{\phi}_{9,10} ...$
Here, $\ket{\phi'} = \mathcal{S}(\ket{\phi})$
is the state $\ket{\phi}$ where the qubits are swapped.
This construction implies that the reduced states 
of $\ket{\tilde \psi}$ on the qubits $1,4,5,8 ...$
are identical, as well as the reduced states
on the qubits $2,3,6,7 ... $
Since the corresponding reduced states are identical, 
the factorizing terms between two qubits (say, $2$ and $3$)
are just squares of some expectation values, hence 
$\vec{v}_1$ and $\vec{v}_2$ are parallel and
$\vec{v}_1 \cdot \vec{v}_2 = \norm{\vec{v}_1} \norm{\vec{v}_2}.$

To perform the second step, note that the state $\ket{\phi}_{k,l}$ 
on the qubits $k,l$ gives rise to some sign distribution of the 
expectation values $x_k, y_k, z_k$ and $x_l, y_l, z_l.$ 
Then we define $\ket{\phi''}$ as follows. We first swap, i.e.
$\ket{\phi''}_{m,n}=\ket{\phi'}_{m,n}= \mathcal{S}(\ket{\phi}_{m,n}),$ 
then, by local unitary transformations, we flip the signs 
of $x_m,z_m$ and $x_n,z_n$  on the qubits $m,n$. Finally, 
we transpose the density matrix of the state, which flips also the
signs of $y_m$ and $y_n.$ Thus we have 
finally $a_k = - a_n$ and $a_l = - a_m$ for $a=x,y,z$ and 
$\ket{\phi''}.$ Note that $\ket{\phi''}$  still saturates
Eq.~(\ref{c2bound}), since the expectation values $a_m a_n$
are not affected.
Then, defining
$
\ket{\psi}= \ket{\phi}_{1,2}\otimes \ket{\phi''}_{3,4} 
\otimes \ket{\phi}_{5,6} \otimes ...
$ 
we arrive at a state for which
$\vec{v}_1 \cdot \vec{v}_2 = -\norm{\vec{v}_1} \norm{\vec{v}_2}.$
Thus, this state saturates Eq.~(\ref{hbound}).

In general, however, the bounds derived by the method above 
are not sharp. Especially, when we consider frustrated lattices, 
the bounds are not sharp, and more sophisticated estimates are 
required. We will discuss one example of a frustrated lattice 
later in detail. Also, if $N$ is not a multiple of $k,$ the 
bound for $k$-producibility may not be sharp. This is, however, 
not a major problem. The energy 
difference between this case and the nearest $N$ which is 
multiple of $k$ is bounded by a constant. Since  we are 
interested in the thermodynamic limit $N \rightarrow \infty, $ 
and the energy difference per bond decreases as $1/N,$ we can 
neglect this case.

What is required to derive similar bounds as Eq.~(\ref{e2bound})
for other spin systems and higher degrees of multipartite 
entanglement? The main ingredient are bounds as in 
Eqs.~(\ref{cibound}, \ref{c1bound}, \ref{c2bound}). 
These bounds depend on the Hamiltonian and on the 
underlying lattice.
For many Hamiltonians and two-qubit blocks, these bounds 
can straightforwardly be computed analytically. But even 
if this is not possible, one can simply compute them by 
numerical minimization over a small number of qubits, 
if desired, this minimizations can be performed with 
assurance of global optimality \cite{lasserre}.

Finally, the reader should note the difference between
the estimation method presented in this Section and the one 
used in Ref.~\cite{us}. The method in  Ref.~\cite{us} does 
not separate between  factorizing and non-factorizing
terms in the Hamiltonian, instead, it estimates the 
complete Hamiltonian via the Cauchy-Schwarz inequality. 
As a consequence, it requires more effort and is restricted 
to one-dimensional systems.

\section{Three simple applications}

In this Section, we will apply the presented method
to various examples of spin systems. We will first
compute energy bounds for $k$-producibility of spin
systems with an anti-ferromagnetic Heisenberg interaction
in various dimensions. Then we will consider the  Ising model
and the XX model in an 
external magnetic field.

\subsection{The Heisenberg model}

Let us first consider the anti-ferromagnetic Heisenberg 
interaction. That is, we consider the Hamiltonian
\be
H_H = \sum_{<i,j>}
X_{i}X_{j}+
Y_{i}Y_{j}+
Z_{i}Z_{j}.
\ee
on a $D$-dimensional lattice. For this model, we can state: 

{\bf Theorem 1.} (a) Let us consider an infinite 
one-dimensional spin system with the Heisenberg interaction. 
Then, the 
energy bounds per bond for one-, two-, 
three-, and four-producible states are given by
\bea
E^{1D}_{1p}  =  -1;
&&
E^{1D}_{2p}  = - \frac{3}{2};
\nonumber \\
E^{1D}_{3p}  = -1.505;
&&
E^{1D}_{4p}  = -1.616.
\eea
The ground state energy per bond is known to be 
$E_0 = - (4 \ln 2 -1) \approx - 1.773 $ \cite{hulthen}, 
thus all the bounds above are violated by the 
ground state.
\\
(b) For the two-dimensional square lattice, the 
respective energies per bond are given by
\bea
E^{2D}_{1p}  =  -1 ;
&&
E^{2D}_{2p}  = - \frac{13}{12} ;
\nonumber \\
E^{2D}_{3p}  = -1.108;
&&
E^{2D}_{4p}  = -1.168.
\eea
Here, the energy per bond in the ground state 
is given by $E_0 = -1.338$ \cite{dowling}.
\\
(c) For the three-dimensional lattice
we have $E_0 = -1.194$ \cite{dowling}
and the thresholds for multipartite 
entanglement read
\bea
E^{3D}_{1p}  =  -1 ;
&&
E^{3D}_{2p}  = - \frac{31}{30} ;
\nonumber \\
E^{3D}_{3p}  = - 1.044;
&&
E^{3D}_{4p}  = - 1.067.
\eea
All the bounds given in this theorem are sharp.

{\it Proof.} The proof of this theorem works just 
as described in the previous section.  For the 
one-producible (i.e. the fully separable) states, 
the bounds have already been shown before \cite{geza, vlatko, dowling}. 
The bounds for two-producible states have been obtained 
analytically (see Lemma 1 in the Appendix); here, the bound 
for the one-dimensional chain was already derived in Ref.~\cite{us}. 
The bounds for the three- and four-qubit case have been 
obtained numerically \cite{mathematica}. Note that for four-producible states and 
$D\geq 2$ several possibilities of four-qubit blocks 
have to be taken into account. The sharpness of the bounds 
follows also as discussed in the previous  section.
$\qed$

\subsection{The Ising model in a transverse magnetic field}

As a second example, let us study the one-dimensional
Ising-model in a transverse magnetic field. That is, we 
consider the Hamiltonian
\be
H_I = \sum_{<i,j>} X_{i}X_{j} + B \sum_{i} Z_{i}.
\ee
The estimation of the energy for $k$-producible states 
can be performed as in the previous section. Only the 
interaction terms with the magnetic field have to be added 
in the definition of $\II(i).$ For instance, for two-qubit blocks, 
we have to compute
\be
C_i = \max_{\ket{\psi}} 
\big[ 
-x_k x_l  - B (z_k + z_l)+\frac{1}{2} (x_k^2 + x_l^2 )\big]. 
\ee
This and similar maximizations can easily be performed 
numerically. The resulting bounds are always sharp. Note 
that the bound for one-producible states has already been 
derived employing a different method in Refs.~\cite{geza,dowling}.

The Ising model is analytically solvable and the thermodynamic 
properties of the thermal states are known \cite{pfeuty}. To 
investigate the multipartite entanglement properties, we  
first compute the energy thresholds $E_{kp}$ for k-producible 
states. We then compare these energies with  the ground state 
energy by calculating the {\it entanglement gap}
\be
E_g (k, B)= E_{kp}(B) - E_0(B),   
\ee
that is the difference between the ground state energy 
and the minimal energy for $k$-producible states 
\cite{dowling}. 
Note that the energy minimum for separable states for a 
quantum Hamiltonian equals the energy minimum of the 
corresponding classical spin chain \cite{geza}.
Thus $E_g(1,B)$ is the energy difference between 
the classical and the quantum Ising spin chains. 
The results are shown in Fig.~3.

\begin{figure}
\centerline{\epsfxsize=0.95\columnwidth
\epsffile{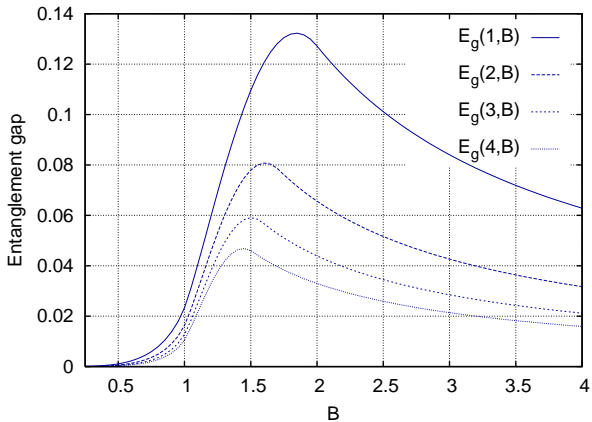} }
\caption{Entanglement gap $E_g (k, B)$ for two- three-, four- and 
five-partite entanglement for the Ising model
in a transverse magnetic field. See text for details.
}
\label{fig_ising}
\end{figure}

To discuss these results, let us consider Fig.~\ref{fig_ising} 
and look at the curve corresponding to $E_g(1,B).$  For a magnetic 
field slightly larger than $B_c=1$ 
the entanglement gap (and thus the entanglement in the thermal 
state) is larger than further from this point. Note that at $B_c=1$ 
the ground state of the Ising model undergoes a quantum phase 
transition. $E_g(n,B)$ for $n>1$ also takes its maximum around $B_c.$
Fig.~\ref{fig_ising} shows that the field corresponding to this 
maximum is decreasing with increasing $n.$

\begin{figure}
\centerline{\epsfxsize=0.95\columnwidth
\epsffile{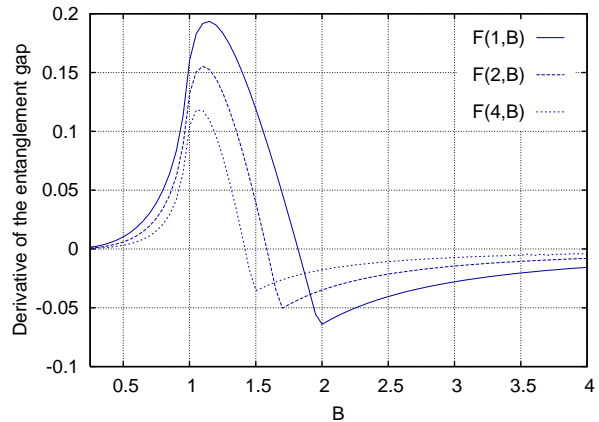} }
\caption{Derivative of the entanglement gap 
$F(k,B)=\partial E_g(k,B)/\partial B$
for $k=1,2$ and $4.$
}
\label{fig_deriv}
\end{figure}

Now let us study $F(k,B)=\partial E_g(k,B)/\partial B$ as the 
derivative of $E_g(k,B)$ with respect  to $B.$ These curves 
are shown in Fig.~\ref{fig_deriv}. On can see directly from
this figure that the slope of $F(1,B)$ has an abrupt change 
at $B_{c,1}=2.$ 
Further analysis shows, that $F(1,B)$ is also non-analytical at 
$B_c = 1.$ The non-analytical point $B_{c}$ corresponds 
to a quantum phase transition of the quantum spin chain, while the 
change in the slope at $B_{c,1}=2$ corresponds to the critical 
point of the classical spin system \cite{DIEGO}.

Similar results can be obtained for $E_g(k,B)$ for $k>1.$
The energy minimum for $k$-producible states equals also 
the energy minimum of a spin model in which blocks of $k$ 
quantum spins interact classically, i.e., in a mean-field 
fashion \cite{TL01}. Again there are non-analytical points
for $F(k,B)$ at $B_c=1$ and at $B_{c,n}>1.$  Fig.~\ref{fig_deriv} 
shows the curves corresponding to $F(1,B)$, $F(2,B)$ and 
$F(4,B).$ It is clearly visible how $B_{c,k}$ approaches 
$B_c=1$ with increasing $n.$ It can also be seen that the 
maximum of $F(k,B)$ also approaches $B_c$ as $k$ increases.
A detailed study of the thermodynamics arising from these 
models intermediate between classical and quantum spin chains 
will be reported elsewhere.

\subsection{The XX model in a magnetic field}
As a third example, we study the one-dimensional XX
model in a magnetic field. The Hamiltonian of this 
model 
is
\be
H_{XX} = \sum_{<i,j>} X_{i}X_{j}+ Y_{i}Y_{j} + B\sum_i Z_{i}.
\ee
The estimation of the energy for $k$-producible states can be 
performed similarly as for the Ising model. Since the XX-model 
can be solved analytically \cite{katsura}, it is now interesting 
to  investigate the regions in the $T$-$B$-plane where multipartite 
entanglement must be present. This has been done in Fig.~5.  Similar 
to the Ising model after the quantum phase transition at $B=2$ the thermal 
states show different forms of multipartite entanglement, 
even at relatively high temperatures.

\begin{figure}
\centerline{\epsfxsize=0.999\columnwidth
\epsffile{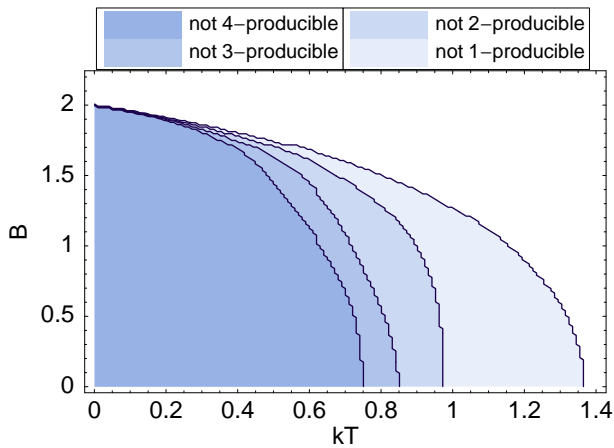} }
\caption{
Entanglement in thermal states of the  XX-model in 
a magnetic field. The regions in the  $T$-$B$-plane
are shown where the different types of multipartite 
entanglement can be detected with our method. 
}
\end{figure}

\section{The Heisenberg model on a triangular lattice}

Let us finally demonstrate with an example that our 
method with some modifications also allows  
the computation of energy thresholds for frustrated 
lattices \cite{frustration}. 
Generally, all lattices can be divided 
into two classes: {\it Bipartite} lattices are lattices, 
where the lattice points can be divided into two 
sublattices, such that each point in each sublattice 
interacts only with points which belong to the other 
sublattice.
An example is the two-dimensional square lattice, 
for which these two lattices form a chessboard-like 
configuration. 
A lattice is called {\it frustrated} if it is not bipartite. 
This terminology 
refers to 
the fact that for such lattices 
the ground state energy per bond is usually larger than 
that for two qubits interacting alone.

Entanglement properties of frustrated systems have also been 
investigated \cite{frusmag, dowling}. Concerning our approach,
the fact that the ground state energy is large makes it difficult 
to derive energy bounds for $k$-producible states which are violated 
by the ground state.

As such an example of a frustrated quantum system we study 
now the Heisenberg model on a two-dimensional triangular 
lattice. That is, we consider the Hamiltonian of 
Eq.~(\ref{XXXmodel}) on the lattice of Fig.~6. Let us shortly 
note some properties of this system. The ground state energy 
per bond in known to be $E_0=-0.726$ \cite{values}. From a 
comparison with a classical spin configuration, it was shown in 
Ref.~\cite{dowling} that the minimal energy per bond 
for fully separable (i.e.~one-producible) states is
$E_{1p}=-0.5.$ Here, we want to derive a bound for two-producible 
states.

If we apply directly the method of the previous section, the 
resulting bound is not violated by the ground state. The reason 
is the following: In the derivation, we used in Eq.~(\ref{csu})
the bound 
$\vec{v}_1 \cdot \vec{v}_2 \geq - \norm{\vec{v}_1} \;\norm{\vec{v}_2}.$
This bound is not sharp for frustrated lattices. 
Thus, we have to make a  more sophisticated estimate.

\begin{figure}
\centerline{\epsfxsize=0.9\columnwidth
\epsffile{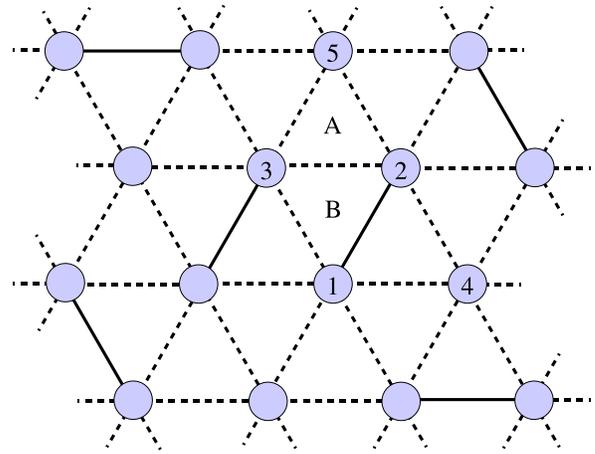} }
\caption{
A two-producible state on a triangular 
lattice. Solid lines represent possible
entanglement between the qubits.
The triangle $(2,5,3)$ is type A, i.e., 
it does not have entanglement
between its qubits. The 
triangle $(1,2,3)$ is of type B since
qubits 1 and 2 may be entangled.
When estimating $C_i$ for the block of the qubits
$1$ and $2,$ the two triangles $(1,2,3)$
and $(1,2,4)$ are estimated via Eq.~(\ref{fb5}), 
since there are definitely of type B.
See text for further details.
}
\end{figure}

First, note that the scalar product $\vec{v}_1 \cdot \vec{v}_2$
represents all factorizing terms in the Hamiltonian. These terms
can be grouped into the contributions corresponding to different
triangles $T_i$. So we can write
\be
2 \vec{v}_1 \cdot \vec{v}_2 = 
\sum_{\rm{triangles}\;\;T_i}
\fh(T_i),
\label{fb1}
\ee
where the triangle contributions $\fh(T_i)$ can be of two types, 
depending on the triangle: For a triangle on the qubits $j,k,l$ 
with no entanglement between the qubits $j,k,l$ (type A triangle, 
see Fig.~6) we have
\be
\fh(T_i)=\sum_{a=x,y,z} 
(a_j \cdot a_k + a_k\cdot a_l + a_l \cdot a_j).
\label{fb2}
\ee
For triangles where two of the three qubits (say, $k$ and $l$)
may be entangled (type B) we have
\be
\fh(T_i)=\sum_{a=x,y,z} (a_j \cdot a_k + a_j\cdot a_l).
\label{fb3}
\ee
The prefactor of two in Eq.~(\ref{fb1}) stems from the fact 
that every bond contributes to two triangles.

Now we need the facts that 
\be
- \Big(
\frac{a_j}{\sqrt{2}}
\cdot 
\frac{a_j}{\sqrt{2}}
+
\frac{a_k}{\sqrt{2}}
\cdot 
\frac{a_k}{\sqrt{2}}
+
\frac{a_l}{\sqrt{2}}
\cdot 
\frac{a_l}{\sqrt{2}}
\Big)
\leq a_j \cdot a_k + a_k\cdot a_l + a_l \cdot a_j,
\label{fb4}
\ee
which holds for all real numbers $a_j,a_k, a_l$, and we need 
the estimate
\be
-
\Big(
\frac{a_k a_l+1}{\sqrt{2}} \cdot \frac{a_j}{\sqrt{2}}
+\frac{a_k a_l+1}{\sqrt{2}} \cdot \frac{a_j}{\sqrt{2}}
\Big)
\leq   a_k \cdot a_j + a_l \cdot a_j.
\label{fb5}
\ee
This estimate holds since $a_i$ and $a_k a_l$ are expectation
values of (tensor products) of Pauli matrices. Namely, from 
the positivity of the  density matrix the relation
$-(1 + a_ka_l) \leq a_k+a_l \leq (1+a_k a_l)$
follows, which results in Eq.~(\ref{fb5}).

The bounds in Eq.~(\ref{fb4}, \ref{fb5}) can be interpreted 
in the following way: for each of the $\fh(T_i)$ there are two 
vectors $\vec{w}^{(1)}_i$ and $\vec{w}^{(2)}_i$ such that
\be
\fh(T_i) \geq - \vec{w}^{(1)}_i \cdot \vec{w}^{(2)}_i.
\ee
If $T_i$ is of the type A then $\vec{w}^{(1)}_i$ has nine 
entries and if $T_i$ is of the type B it has six entries. 
Note that the definition of $\vec{w}^{(1)}_i$ etc. does not 
directly correspond to the $\RR$ and $\NN$ in the definition 
of $\vec{v}_{1/2}.$

Thus, if we define
\be
\vec{W}^{(1)} = \bigoplus_i \vec{w}^{(1)}_i,\;\;\;
\vec{W}^{(2)} = \bigoplus_i \vec{w}^{(1)}_i,
\label{fb6}
\ee
it follows that $2 \vec{v}_1 \cdot \vec{v}_2  \geq 
- \vec{W}^{(1)} \cdot \vec{W}^{(2)}.$  Using the 
Cauchy-Schwarz inequality yields
\be
\vec{v}_1 \cdot \vec{v}_2 
\geq 
-\frac{1}{4} 
\big( \norm{\vec{W}^{(1)}}^2 + \norm{\vec{W}^{(2)}}^2
\big).
\label{fb7}
\ee

The right hand side of Eq.~(\ref{fb7}) is a sum of many 
squares of expectation values of one- or two qubit 
observables. Each of these expectation values 
originates
from a certain block. By counting carefully the 
contributions of each block, we can now estimate them 
separately for each block.

A one-qubit block on the qubit $k$, contributes to the 
estimates of six triangles. These may be triangles of 
type A or B. An estimate of one triangle of the type A
results in two times a contribution $(a_k/\sqrt{2})^2$
in $\norm{\vec{W}^{(1)}}^2 + \norm{\vec{W}^{(2)}}^2,$ 
one in $\norm{\vec{W}^{(1)}}^2$ and one in 
$\norm{\vec{W}^{(2)}}^2$ (see Eq.~(\ref{fb4})). 
A triangle of the type B gives also two times this 
contribution, but now either two times in 
$\norm{\vec{W}^{(1)}}^2$ or two times in 
$\norm{\vec{W}^{(2)}}^2$ (see Eq.~(\ref{fb5})).
Thus, in analogy to Eqs.~(\ref{cibound}, \ref{c1bound}), we have 
to estimate
\be
C_i = 
\max_{\ket{\psi}} 
\big[\frac{1}{4} \cdot 6 \cdot 2
(\frac{x_k^2}{2}+ \frac{y_k^2}{2}+ \frac{z_k^2}{2}) \big] 
= \frac{3}{2}.
\ee

A two-qubit block on the qubits $k$ and $l$ contributes to 
ten triangles. Two of them contain both the qubits $k$ and $l$
and are thus of the type B (see Fig.~6). Each of them
contribute two times $(a_ka_l+1/\sqrt{2})^2$ to 
$\norm{\vec{W}^{(1)}}^2 + \norm{\vec{W}^{(2)}}^2,$ 
either both in $\norm{\vec{W}^{(1)}}^2$ or both in 
$\norm{\vec{W}^{(2)}}^2$ (see Eq.~(\ref{fb5})). For the other 
eight triangles, it does not matter as for the one-qubit blocks
whether they are of the type A or B. 
Thus, we have to estimate
\bea
C_i &=& \max_{\ket{\psi}} 
\Big[ -x_k x_l - y_k y_l - z_k z_l +
\nonumber 
\\
&+&
\frac{2 \cdot 2 }{4}
\big(
\frac{(1+x_k x_l)^2}{2}+\frac{(1+y_k y_l)^2}{2}
+\frac{(1+z_k z_l)^2}{2}
\big)
\nonumber 
\\
&+&
\frac{4 \cdot 2}{4}
(\frac{x_k^2}{2}+ \frac{y_k^2}{2}+ \frac{z_k^2}{2}
+\frac{x_l^2}{2}+ \frac{y_l^2}{2}+ \frac{z_l^2}{2})
\Big] 
= 4.
\nonumber 
\\
\label{fb8}
\eea
This follows similarly as before, see also Lemma 2(b) in 
the Appendix.

With these bounds we have for the two-producible state
$\mean{H}\geq -\sum_i C_i,$ which finally results in
a minimal energy per bond of
\be
E_{2p} = - \frac{2}{3},
\label{fb9}
\ee
which is clearly violated by the ground state. 
Thus, at a considerable temperature, in a frustrated 
triangular Heisenberg lattice the thermal state is 
not two-producible  and the effects of multipartite 
entanglement cannot be neglected.

Finally, it should be noted that it is not clear whether
the bound in Eq.~(\ref{fb9}) is sharp. Two facts suggest
that this is not the case. First, the use of the Cauchy-Schwarz
inequality in Eq.~(\ref{fb7}) is probably not sharp, since
all the vectors $\vec{W}^{(1)}$ and  $\vec{W}^{(2)}$ are usually
not parallel. Second, the maximum in  Eq.~(\ref{fb8}) is obtained 
for a separable state, no entanglement is required to saturate 
this bound.

\section{Conclusion}

In conclusion, we have developed a method to investigate the 
presence of multipartite entanglement in spin models. Our method
relies on energy bounds for certain classes of multipartite 
entangled  states and can be applied to any state, not only 
to thermal states. We discussed different examples and showed
that our ideas can in principle also be applied to the investigation
of frustrated systems.

The results of this paper may be applied in several directions. 
On the one hand, based on our energy thresholds one may derive 
temperature bounds below which entanglement must be present. 
Then, one may try to design methods to extract this entanglement 
and make it useful for some tasks. 

On the other hand, our results can also 
be used to gain theoretical insight concerning the validity of 
ground state approximations. For example, one might be interested 
in the ground state energy $E_0.$ To approximate this, one may consider 
some mean-field like approximation, where the trial wave function is 
a product state with respect to all qubits \cite{richert}. This corresponds 
to a one-producible state. Concerning the justification of this 
approximation, we can say with our method the following: If $E_{1p} 
\approx  E_{2p}$ the approximation might be justified, while if 
$E_{1p} \gg  E_{2p}$ this is clearly not the case.

\section{Acknowledgment}

We would like to thank H.J.~Briegel, J.I.~Cirac, 
M. Dowling, W.~Dür, O.~Gittsovich,  A.~Hamma, 
P.~Hyllus, D. Porras, J.~Richert, E.~Rico and 
T.~Roscilde for useful discussions. 
We also acknowledge the support of the 
EU (IST-2001-38877, OLAQUI, ProSecCo, QUPRODIS, RESQ, SCALA)
the FWF, the DFG and the Kompetenznetzwerk 
Quanteninformationsverarbeitung der Bayerischen
Staatsregierung. 
G.T. thanks the support of the European Union (Grant Nos.
MEIF-CT-2003-500183 and MERG-CT-2005-029146) and
National Research Fund of Hungary  OTKA under Contract No.
T049234.

\section{Appendix}

Here, we prove two useful bounds for our proofs. 

{\bf Lemma 1.} For two-qubit state $\ket{\psi}$ on the qubits
$k,l$  and $\gamma \geq 1/2$ we have the following sharp bound: 
\bea
&&-x_k x_l-y_k y_l-z_k z_l + 
\gamma (x_k^2 + y_k^2 + z_k^2 + x_l^2 + y_l^2 + z_l^2)
\nonumber 
\\
&&
\;\;\;\;\;\;
\leq
1+2\gamma + \frac{1}{2\gamma}.
\label{bound1}
\eea
{\it Proof.} We have to maximize the left hand side of
the inequality over all states. First note that a generic 
quantum state can be written as 
\be
\ketbra{\psi}
=\frac{1}{4}\sum_{i,j=1,x,y,z}\lambda_{ij} \sigma_i \otimes \sigma_j,
\ee
with $\sigma_1=\eins.$ Here, 
$\lambda_{ij}=\bra{\psi}\sigma_i \otimes \sigma_j\ket{\psi}$ 
holds, thus we can directly maximize over all allowed 
$\lambda_{ij}.$ Let us first consider local unitary 
transformations. These transformations act on 
$\lambda_{ij}$ as 
$
(\lambda_{ij}) \rightarrow (1\oplus O_L) (\lambda_{ij}) 
(1\oplus O_R)
$
where $O_L$ and $O_R$ are orthogonal $3\times 3$ matrices. 
Here, $1\oplus O_R$ denotes a $4 \times 4$ matrix with a block
structure, i.e., with ``$1$'' in the left upper corner and $O_R$ denotes
the $3 \times 3$ block in the right bottom corner. 
These transformations do not change the purities of the 
reduced states, thus they do not change $x_k^2 + y_k^2 + z_k^2$ 
and $ x_l^2 + y_l^2 + z_l^2.$ Furthermore, 
$|x_k x_l| + |y_k y_l| + |z_k z_l| $  is the sum of 
the absolute values of
the diagonal elements of the $3 \times 3$ submatrix 
$\lambda^{red}_{ij}=\lambda_{i,j=x,y,z}$. This sum is 
maximized when $\lambda^{red}$ is brought 
to diagonal form via a singular value decomposition. 
This decomposition can be performed by local unitary 
operations. 

Thus it suffices to consider 
$\ket{\psi}=\alpha \ket{00}+\beta \ket{11}$
with $\alpha^2+\beta^2=1,$ since for that state we have
\be
(\lambda_{ij})
=
\left( 
\begin{array}{cccc}
1&0&0&{\alpha^2-\beta^2}\\
0&2{\alpha\beta}&0 & 0 \\
0&0&-2{\alpha\beta} & 0\\
{\alpha^2-\beta^2} &0&0&{1} 
\end{array}
\right). 
\label{lambdamatrix}
\ee
The final maximization over all $\alpha$ can then be 
directly performed. The left hand side of Eq.~(\ref{bound1})
is maximized for 
$4\alpha^2=2-\sqrt{4\gamma^2-1}/\gamma$
which proves the bound in Eq.~(\ref{bound1}).
$\qed$
\\
\\
\indent
{\bf Lemma 2.} (a) For two-qubit state $\ket{\psi}$ on the 
qubits $k,l$  and $\gamma \geq 1/2$ we have the 
following sharp bound: 
\be
- x_k x_l - y_k y_l + 
\gamma (x_k^2 + y_k^2 + x_l^2 + y_l^2)
\leq
1+2\gamma + \frac{1}{8\gamma}.
\label{bound2}
\ee
(b) Similarly, we have
\bea 
&&\frac{3}{2} + \frac{1}{2} (x_k x_l^2 + y_k y_l^2+z_k z_l^2) +
\nonumber
\\
&& +(x_k^2 + y_k^2 + z_k^2 + x_l^2 + y_l^2 +z_l^2)
\leq
4.
\label{bound3}
\eea

{\it Proof.} (a) For the minimization it does not 
matter whether we minimize over the observables $X,Y$ 
[as in Eq.~(\ref{bound2})] or $X,Z.$ Then, the bound 
can be derived as in the proof of Lemma 1. One arrives 
again at Eq.~(\ref{lambdamatrix}), now only one of the terms
$2\alpha \beta$ has to be omitted in the final maximization. 
(b) This can also be derived as in Lemma 1. When applying
local unitary transformations, $((x_k x_l)^2 + (y_k y_l)^2+(z_k z_l)^2)$
is maximal, when the $3\times 3$ matrix is diagonal. Finally,  one has only 
to maximize over $\alpha$ again.
$\qed$


\begin{thebibliography}{99}
\bibitem{bemerkung} Note that this is also true for the 
classification concerning $k$-separability or the 
classification of $W$/$GHZ$ states for three qubits.
\bibitem{remark1}
In our example, one may also prove this explicitly and 
take the two-qubit state 
$\ketbra{\phi}= 
(
\eins \otimes \eins + 
2a\sqrt{1-a^2}
(\sigma_x \otimes \sigma_x +\sigma_y \otimes \sigma_y)
-\sigma_z \otimes \sigma_z
-(2a^2-1)(\sigma_z \otimes \eins - \eins\otimes \sigma_z)
)/4$
with $a=-\sqrt{1/2-\sqrt{2}/3},$  which saturates the bound 
in Eq.~(\ref{c2bound}) to find the total state saturating
Eq.~(\ref{e2bound}).
\bibitem{mathematica} For all our numerical computations
we used the optimization packages of the computer algebra 
system ``Mathematica''.
\end{thebibliography}
\end{document}